\documentclass[reprint,amsmath,amssymb,pra,superscriptaddress,showpacs]{revtex4-1}
\usepackage{braket,graphicx,epsfig}
\usepackage[utf8x]{inputenc}
\usepackage{subfigure}
\usepackage[dvips]{color} 

\newcommand{\gl}{\lambda}
\newcommand{\gs}{\sigma}
\newcommand{\upd}{\textrm{d}}
\definecolor{violeta}{cmyk}{0.07,0.90,0,0.34}

\begin{document}
\title{A statistical portrait of the entanglement decay of two-qubit memories}

\author{Karen M. Fonseca-Romero}
\email{kmfonsecar@unal.edu.co}
\affiliation{Universidad Nacional de Colombia - Bogot\'a, Facultad de Ciencias,  Departamento de F{\'isica}, Carrera 30 Calle 45-03, C.P. 111321, Bogot{\'a}, Colombia}

\author{Juli{\'a}n Mart{\'i}nez-Rinc{\'o}n}
\affiliation{Universidad Nacional de Colombia - Bogot\'a, Facultad de Ciencias,  Departamento de F{\'isica}, Carrera 30 Calle 45-03, C.P. 111321, Bogot{\'a}, Colombia}
\affiliation{Department of Physics and Astronomy, University of Rochester, Rochester, New York 14627, USA}

\author{Carlos Viviescas}
\affiliation{Universidad Nacional de Colombia - Bogot\'a, Facultad de Ciencias,  Departamento de F{\'isica}, Carrera 30 Calle 45-03, C.P. 111321, Bogot{\'a}, Colombia}

\date{\today}

\begin{abstract}
We present a novel approach to the study of entanglement decay, which focuses on collective properties. 
As an example, we investigate the entanglement decay of a two-qubit system, produced by local identical reservoirs acting on the qubits, for three experimentally and theoretically relevant cases.
We study the probability distributions of disentanglement times, a quantity independent of the measure used to quantify entanglement, and the time-dependent probability distribution of concurrence. Analytical results are obtained for initially uniformly distributed pure states.
The calculation of these probability distributions gives a complete insight on how different decoherence channels affect the entanglement initially contained in the set of two-qubit pure states. 
Numerical results are reported for randomly distributed initial mixed states.
Although the paper focuses in Markovian noisy channels, we show that our results also describe non-autonomous and non-Markovian channels.
\end{abstract}


\maketitle

\section{Introduction.}
Entanglement is a key resource for quantum information processing and communication, which can be manipulated, broadcasted, controlled and distributed \cite{Horodecki2009}. 
A better understanding of the effects of decoherence on entanglement could accelerate the development of technological applications of quantum information protocols. 
Its study under realistic, dissipative conditions, is of fundamental importance to assess the resilience of quantum information processing. 
The dynamics of quantum coherence and entanglement can be strikingly different. Under the influence of a noisy environment, coherences usually decay asymptotically in time, while entanglement can do so in a finite time \cite{Zycz01b,ReviewEberly09}.
Extensive work \cite{Eberly03,YuEberly2004PRL93a140404,Jakob04,Eberly06_2env,Eberly06ClassicalNoise,Jakob04,FiniteT08,Rau09,Lopez08}, including several experimental demonstrations \cite{ExpDavidovich07,Farias09,Barr10}, has been devoted to this remarkable phenomenon, sometimes called entanglement sudden death (ESD). A complex, yet partial picture of the entanglement dynamics, has emerged.
Recently, a complete characterization of the dynamics of a two-qubit system in which only one of the qubits is coupled to a noisy channel \cite{Konr08} has been presented and general approaches, based on geometrical arguments \cite{Geometry07} and quantum trajectories \cite{cv}, have been proposed.

In this work, adopting a broader perspective, we render a comprehensive description of the entanglement dynamics in two-qubit systems when corrupted by environment-induced decoherence. Our approach relies on a rather novel tool in the area: the statistical distributions of concurrence and of the ESD  times. Obtaining analytic expressions for them is far from being a symple task. In fact, previous analytic work was restricted to averages \cite{Poyatos1997} and often relied on approximations \cite{Fonseca2005PRL95a140502}. Other reported explorations of entanglement evolution in bipartite \cite{Zycz01b}  and multipartite systems \cite{Tier10}, similar in spirit to our approach, are both numerical. Our analytic formulation, employed before in a kinematical context \cite{Martinez2009RCF41p524}, allows us to depict a highly detailed image of the entanglement time evolution. 

This paper is organized as follows. The dynamical evolution of a two-qubit system coupled to local channels is introduced in section \ref{sec:xx}, where the probability densities of the disentanglement times and of concurrence are defined. We consider three models of decoherence for a single qubit and assume identical environments for each qubit. In the next four sections the results for the probability densities of the disentanglement times and of concurrence are  evaluated for the chosen environments, assuming initial pure states uniformly distributed. The results for mixed states, a single noisy channel, and non-autonomous systems are briefly presented in sections \ref{sec:mixedstates}, \ref{sec:singlechannel} and \ref{sec:nonautonomous}, respectively. Section \ref{sec:nonmarkovian} is devoted to the demonstration that, although we concentrate on Markovian processes with constant decay rates, our results also apply to non-Markovian processes. Some conclusions are drawn in the last section of this paper.

\section{Disentanglement probability distributions}\label{sec:xx}
We investigate the \emph{global} entanglement dynamics of two-qubit systems coupled to local reservoirs (channels), a situation likely to hold for spatially separated qubits. The qubits are assumed to be quantum memories; they do not follow independent unitary dynamics nor do they interact with each other. Whenever the initial system-environment state is separable with vanishing quantum discord \cite{OllivierPRL88a017901,Henderson2001JPA34p6899}, the dynamical evolution of the state of the two-qubit system is given by a completely positive map \cite{Rodriguez2008JPA41p205301,Shabani2009PRL102p100402}. Then, $\rho(t)$, the system state at the physical time $t$, is given by
$\rho(t) = \sum_{ij} \left(E_i^{(1)} \otimes E_j^{(2)}\right) \rho(0) \left(E_i^{(1)\dagger} \otimes E_j^{(2)\dagger}\right)=\Lambda_t(\rho(0)),$
where the time-dependent Kraus operators $E_i^{(a)}=E_i^{(a)}(t)$, acting on the $a$-th qubit, satisfy the trace-preserving condition $\sum_{i} E_i^{(a)\dagger} E_i^{(a)} = \mathbb{I}^{(a)}$, $ a=1,2$. These operators can be experimentally determined by a quantum process tomography \cite{Mohseni2008PRA77a032322}. 

We chose pairs of identical environments to act on the qubits, and regard the experimentally relevant cases of depolarizing (D), amplitude-damping (AD), and phase-damping (PD) channels. These generally non-Markovian channels are valid for finite or infinite environments and weak or strong coupling \cite{FonsecaRomero2012}. 
The Kraus operators for these noisy environments, in terms of the (reparameterized) time $q$, are ($0\leq q\leq 1$):  $E_0(q) = \sqrt{1-{3q}/{4}}\, \mathbb{I}$ and $E_i(q) = \sqrt{{q}/{4}}\, \gs_i$, $i=1,2,3,$ for case D; $E_0 = \ket{0}\!\bra{0}+\sqrt{1-q}\ket{1}\!\bra{1}$ and $E_1 =\sqrt{q}\ket{0}\!\bra{1}$ for case AD; and $E_0 = \sqrt{1-q}\,\mathbb{I}$, $E_1 = \sqrt{q} \ket{0}\bra{0}$, and $E_2 = \sqrt{q} \ket{1}\bra{1}$ for case PD. Here $\gs_1,\gs_2$, and $\gs_3$ denote, as usual, the three Pauli matrices. 
Unless otherwise stated, we focus on the Markovian scenario, with constant decay rates, when $q(t)=1-\exp(-\gamma t)$, $\gamma$ being a positive constant.
The motivation behind our selection of channels lies in their qualitative different long-time behavior, $q\rightarrow 1$, determining the dynamical evolution of entanglement on a gross level \cite{Geometry07}. All states will suffer ESD if the asymptotic state belongs to the interior of the set of separable states $\mathcal S$ (case D). Some states will separate asymptotically if the asymptotic state belongs to $\partial \mathcal S$, the border between $\mathcal S$ and $E$, the set of entangled states (case AD). The case PD displays a set of stationary separable states, diagonal in the standard basis, which includes 4 pure states in $\partial \mathcal S$.

Rather than compute the concurrence (or any other entanglement measure) for specific initial conditions, we pursue a statistical approach to entanglement dynamics in open systems. Although randomly chosen normalized pure states $\ket{\psi} = \psi_{00}\ket{00}+ \psi_{01}\ket{01}+ \psi_{10}\ket{10}+ \psi_{11}\ket{11}$ are used for our calculations, other choices might be sensible. We choose pure states because they generally offer higher quantum correlations than mixed states, a key point for quantum information protocols. Moreover, they are a convenient idealization of the almost pure states that can be produced under experimental conditions. 
If minimal prior knowledge about the state of the system is assumed \cite{Hall}, or invariance under unitary transformations is required, it becomes natural to use the uniform (Haar) measure for all possible pure states of the system. 
Hence, the probability to find a state in a small volume $d^2\psi=\prod_{ij}d\psi_{ij}\,d\psi_{ij}^*$ around $\psi$ is  
$p(\psi)d^2\psi= \Gamma(4)\delta(1 - \braket{\psi|\psi} \,)/\pi^4 d^2\psi$, where $\Gamma(x)$ denotes, as usual, the Gamma function.

We characterize the disentanglement process through the reparameterized separation time probability density,
\begin{equation}
\label{eq:pts}
p(q_S) = \int \! d^2 \psi \, \delta(q_S-q_S(\psi))p(\psi) \,,
\end{equation}
where $q_S(\psi)$ is the ESD time, i.e., the time at which $\rho(q)=\Lambda_q(\ket{\psi}\bra{\psi})$, becomes separable.
This distribution is independent of the measure used to quantify the entanglement of the system.  
However, $p(q_S)$ may overestimate entanglement if many states sustain small values of entanglement for long times, before undergoing ESD. Therefore, a complete characterization of entanglement decay also requires its quantification. 

The global evolution of entanglement is characterized by using the concurrence probability density,
\begin{equation}
\label{eq:PCt}
p(C;q) = \int \! d^2 \psi \, \delta(C-C(q;\psi))p(\psi) \,,
\end{equation}
where $C(q;\psi)$ is the Wootter's concurrence \cite{Wootters1998a} of the state $\rho(q)=\Lambda_q(\ket{\psi}\bra{\psi})$.  
Concurrence is given by 
$C(q;\psi) \equiv \max\{0, \sqrt{\gl_1}- \sum_{i=2}^4 \sqrt{\gl_i}\}$,
where $\gl_i$ are the eigenvalues of the matrix $\rho(q) \left(\gs_2^{(1)}\otimes\gs_2^{(2)}\right) \rho(q)^* \left(\gs_2^{(1)}\otimes\gs_2^{(2)}\right)$, being $\gl_1$ the largest one. Here $\rho^*$ is the complex conjugate of $\rho$ in the standard basis, and $\gs_2$ is the second of Pauli matrices.
Then, for our chosen ensemble of initial states, $p(C;q)dC$ is the probability to find a state with concurrence value in an interval $dC$ around $C$ at time $q$. In particular, at the initial time $q=0$, eq. \eqref{eq:PCt} provides the concurrence distribution over an ensemble of uniformly distributed pure states, $p(C;0) = 3C\sqrt{1-C^2}$ \cite{Karol,Martinez2009RCF41p524}. 
The contribution of the states with vanishing concurrence to the distribution $p(C;q)$ can be singled out by writing $p(C;q)=\tilde{p}(C;q) + S(q)\delta(C)$, where $\tilde{p}(C;q)$ is the probability density for strictly positive concurrence and $S(q)=1-\int_0^{C_M}\!dC \, \tilde{p}(C;q)$ is the probability to find a separable state at a time $q$, where $C_M=C_M(q)$ is the maximum concurrence at that time.

We will show that ${p}(C;q)$ and $p(q_S)$ provide a complete portrait of the entanglement behavior for all times, enabling us to monitor entanglement degradation, concentration, and even loss. Despite the dissimilar dynamics of the D, AD, and PD channels, in all cases we find analytical expressions for the disentanglement time and the concurrence, for initial pure states, and use them to evaluate the probability distributions of separation times $p(q_s)$ and of concurrence $p(C;q)$ as a function of the reparameterized time. Our results were checked using purely numerical methods. 

\section{Probability distributions of the disentanglement times.}
The disentanglement times, 
\begin{eqnarray}
 q_S^{\footnotesize{\textrm{(D)}}} &=&1-1/ \sqrt{1+2C_0},\\ 
 q_{S}^{\footnotesize{\textrm{(AD)}}}&=&C_0/ 2|\psi_{11}|^2, \\
 q_S^{\footnotesize{\textrm{(PD)}}} &=& 1-\left(\sqrt{C_0^2+d^2}-C_0\right)/d
 \end{eqnarray}
for identical depolarizing, amplitude-damping and phase-damping channels, respectively, are found from the condition of separability, $C(q_S,\psi)=0$.
Here, we have set $d=4|\psi_{00}\psi_{01}\psi_{10}\psi_{11}|^{1/2}$, and $C_0=2|\psi_{00}\psi_{11}-\psi_{01}\psi_{10}|=C(0,\psi)$, and have used superscripts indicating the case/channel we are considering. We use Eq. \eqref{eq:pts} and these expressions for the separation time to obtain the following probability distributions of disentanglement time, 
\begin{equation*}
p^{\footnotesize{\textrm{(D)}}}(q_S)=\frac{3 q_S(2-q_S)  \sqrt{\left(q_S^2-2 q_S+2\right) \left(3 q_S^2-6 q_S+2\right)}}{4 (1-q_S)^7},
\end{equation*} 
\begin{equation*}
p^{\footnotesize{\textrm{(AD)}}}(q_S)=
\frac{q_S^2-1}{2\,q_S\,(1+q_S^2)^2}+\frac{\arctan\left(q_S\right)}{2\,q_S^2}+\frac{2+\pi}{8}\delta(q_S-1),
\end{equation*} 
\begin{equation*}
p^{\footnotesize{\textrm{(PD)}}} (q_S)
= \int d\theta\, ds\, dr\, P_J(s,r) \delta(q_S-q^{\footnotesize{\textrm{(PD)}}}_S)/2\pi.
\end{equation*} 
In the latter case the dependency on the angle $\theta$ comes from the initial concurrence
$C_0=\sqrt{2}\sqrt{s^2+r^2+(s^2-r^2)\cos(\theta)}$.
Here, $P_J\left(s,r\right)= 
\frac{24 (s^2-r^2)}{\sqrt{1-4r^2}} K\left( \frac{1-4s^2}{1-4 r^2} \right)$ stands for the joint probability distribution of $r$ and $s$, with $s+r=2|\psi_{00}\psi_{11}|$ and $s-r= 2|\psi_{01}\psi_{10}|$.  The complete elliptic integral of the first kind and elliptic modulus $k$ is denoted by $K(k^2)$. 
\begin{figure}[h]
 \centering
 \includegraphics[width=6cm,angle=-90]{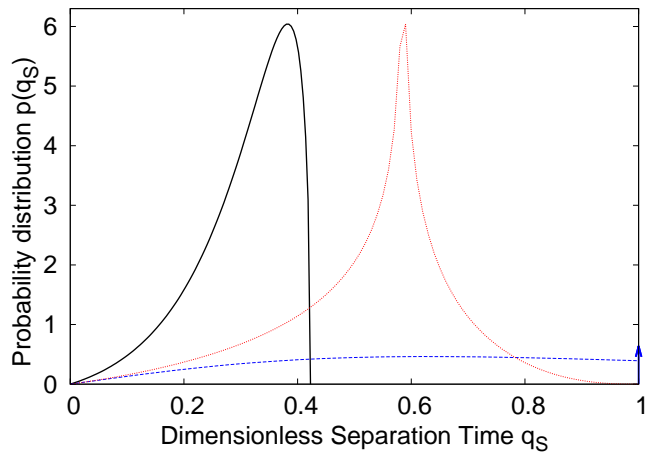}
 \caption{(Color online) Disentanglement-time probability distributions: solid line (black) for local depolarizing channels (D); dashed line (blue), including a delta function contribution, at $q_s=1$, for local amplitude damping channels (AD); and dotted line (red) for local phase damping channels (PD).}
 \label{fig:TimeProbabilities}
\end{figure}

The probability distributions of the separation times (Fig. \ref{fig:TimeProbabilities}) display completely different behaviors. All present
a rather conspicuous peak at around $q_S=0.38$, $q_S=1$, and $q_S=0.59$ for cases D, AD, and PD, respectively. However, the first is the widest, and the second is the more narrow and is actually a delta distribution. While in the first case all states become separable before $q_S=1-1/\sqrt{3}\approx0.42$, only $\frac{6-\pi}{8}\approx35.73\%$ of the states experience ESD in the second case, and a zero-measure set of states separate asymptotically in the third case. The three cases we have considered can be tell apart by their disentanglement time probability distributions, and the fraction of entangled states at a given time can be calculated. However, it is not clear how much entanglement remains at a given time. This is especially important in the AD case, because the entanglement of most states ($64.27$\%) decay asymptotically. In the next three sections, we consider the dynamics of the concurrence probability distribution for each case.

\section{Depolarizing channels.} 
\begin{figure}[h]
 \centering
 \includegraphics[width=6cm,angle=-90]{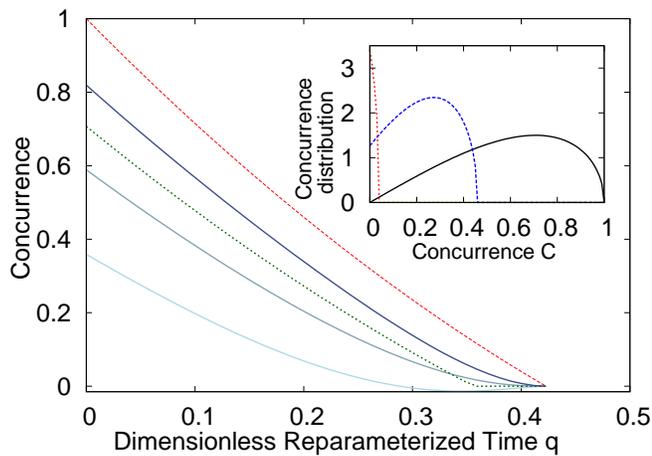}
 \caption{(Color online) Identical but independent depolarizing channels. The solid blue lines represent the average concurrence plus a standard deviation (darkest line), the average concurrence, and the average concurrence minus a standard deviation (lightest line); maximum concurrence is plotted in red (dashed line), and the value of concurrence at which the distribution peaks is plotted in green (dotted line). Inset: From right to left, concurrence distributions for $q=0,0.2$ and $0.4$.}
 \label{fig:D}
\end{figure}

We start by investigating the case in which each qubit is coupled to a depolarizing channel. As time goes by, the depolarizing channels induce decoherence on the two qubits and their initial entanglement is lost. The concurrence at time $q$,
\begin{equation}
\label{eq:ConcurrenceD}
C^{\footnotesize{\textrm{(D)}}}(q;C_0)=\max\left\{0,C_0(1-q)^2-\frac{1}{2} q(2-q)\right\} \,,
\end{equation}
depends only on (time and) the concurrence of the initial state $C_0$. The loss of entanglement in the system is uniform, and initial maximally entangled states remain as the states with the highest concurrence for all times, $C_{M}(q) = 1-\frac{3}{2}q(2-q)$. For an ensemble of uniformly distributed initial pure states, the concurrence probability density at a time $q$
\begin{equation*}
\tilde{p}^{\footnotesize{\textrm{(D)}}}(C;q) = 3 \frac{C+\frac{1}{2}q(2-q)}{(1-q)^4} \sqrt{1-\left(\frac{C+\frac{1}{2}q(2-q)}{(1-q)^2}\right)^2}\,,
\end{equation*}
follows from Eq. \eqref{eq:ConcurrenceD} and $p(C;0)=3C\sqrt{1-C^2}$.
In the inset of Fig. \ref{fig:D} we draw $\tilde{p}^D(C;q)$ for different values of time, $q$. 

In Fig. \ref{fig:D}, entanglement degradation is apparent. Indeed, due to the loss of entanglement, the whole distribution moves towards zero and the average concurrence $\overline{C_q}=\int_0^{C_M} dC\, C\,\tilde{p}^{\footnotesize{\textrm{(D)}}}(C;q)$ decreases. At the same time entanglement concentrates around its mean value because the standard deviation $(\overline{C_q^2} - \overline{C_q}^2)^{1/2}$ falls as $q$ increases. Finally, entanglement is lost: the probability to obtain an entangled state $\int_0^{C_M}\! dC\, \tilde{p}^{\footnotesize{\textrm{(D)}}}(C;q)$ declines with time. 

Entanglement degradation, concentration and loss are displayed in all the examples considered in this work, and seem typical for open markovian systems,  
whose asymptotic states belong to $\mathcal S$ and $\partial \mathcal S$. This asymptotic behavior originates also entanglement concentration due to the shrinking average distance between the states that remain entangled. Entanglement is lost when states cross the border between entangled and separable states. Thus, the concurrence probability distribution provides a  complete picture of the process of entanglement decay. However, the most salient features of this process are contained in the evolution of the concurrence maximum, its average, its deviation and the separation-time probability distribution.


\section{Amplitude-damping channels.} 
In our second example each qubit experiences dissipation under the action of an amplitude-damping channel. 
For an initial two-qubit pure state, the concurrence evolves in time as \cite{cv}
\begin{equation}
C^{\footnotesize{\textrm{(AD)}}}(q;C_0,\psi_{11})=\max\left\{0,(1-q)\left(C_0 - 2 |\psi_{11}|^2 q\right)\right\} \,.
\end{equation}
We notice that the set of initially entangled states separates into two classes: those states for which $C_0<2|\psi_{11}|^2$, becoming separable at a finite time; while all the others states, for which $C_0>2|\psi_{11}|^2$, reaching separability only asymptotically. Hence, contrary to the previous case, the loss of concurrence is not uniform: when comparing states with the same initial concurrence, those with smaller $|\psi_{11}|$ are more robust. At a given time $q$, the maximum concurrence $C_{M}=1-q$ corresponds to initial states of the form $(\ket{01} + \exp(i\phi)\ket{10})/\sqrt{2}$, with arbitrary phase $\phi$.
\begin{figure}[h]
 \centering
 \includegraphics[angle=-90,width=8.5cm]{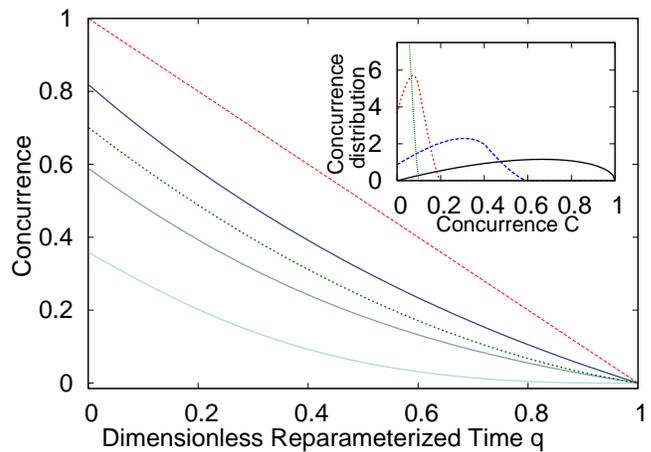}
 \caption{(Color online) Identical but independent amplitude-damping channels. The solid blue lines represent the average concurrence plus a standard deviation (darkest line), the average concurrence, and the average concurrence minus a standard deviation (lightest line); maximum concurrence is plotted in red (dashed line), and the value of concurrence at which the distribution peaks is plotted in green (dotted line). Inset: From right to left, concurrence distributions for $q=0, 0.4, 0.8$ and $0.9$.}
 \label{fig:AD}
\end{figure}
Since $C^{\footnotesize{\textrm{(AD)}}}(q)$ depends on both the initial concurrence and the state component in the $\ket{11}$ direction, the joint probability distribution $P_J(|\psi_{11}|^2,C_0)$ over all two-qubit pure states is calculated first in order to evaluate the concurrence distribution $\tilde{p}^{\footnotesize{\textrm{(AD)}}}(C)$. This joint probability either vanish or is equal to $3\,C_0\log ( (1+\sqrt{1-C_0^2})/z)$, where $z=\max(1-\sqrt{1-C_0^2},\,2\,|\psi_{11}|^2)$ for $ 2|\psi_{11}|^2\leq 1+\sqrt{1-C_0^2}$. 
The concurrence probability distribution $\tilde{p}^{\footnotesize{\textrm{(AD)}}}(C;q)= \int d{|\psi_{11}|^2\,}d{C_0}\, \delta(f)  P_J (|\psi_{11}|^2,C_0)$ (where $f={C}-C^{\footnotesize{\textrm{(AD)}}}(q;C_0,\psi_{11})$)  can be calculated analytically, but the expression is cumbersome and not illuminating. 

The concurrence probability distribution is depicted in the inset of Figure \ref{fig:AD}, for some values of time $q$. As in the previous case the degradation of concurrence (the disentanglement) is more important for smaller (larger) times. The separation between these processes is more conspicuous here because most states disentangle asymptotically. The asymptotic massive disentanglement is not evident in the graphic of average concurrence.

\section{Phase-damping channels.} 
\begin{figure}[h]
 \centering
 \includegraphics[angle=-90,width=8.5cm]{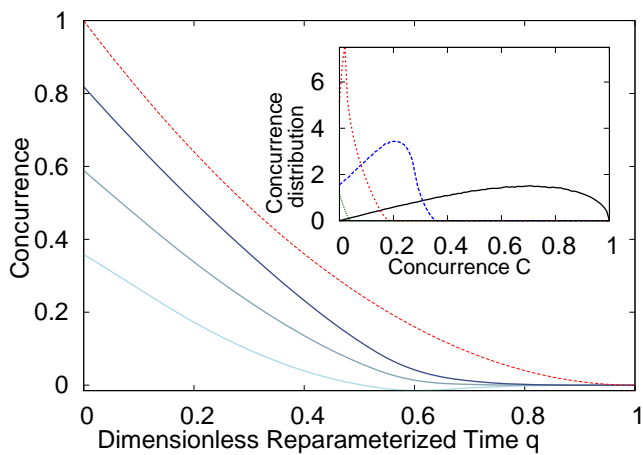}
 \caption{(Color online) Identical but independent phase-damping channels. The solid blue lines represent the average concurrence plus a standard deviation (darkest line), the average concurrence, and the average concurrence minus a standard deviation (lightest line); maximum concurrence is plotted in red (dashed line). Inset: From right to left, concurrence distributions for $q=0,0.4,0.57$ and $0.8$.}
 \label{fig:PD}
\end{figure}

In our last example each qubit evolves under the action of a dephasing channel, which washes away the coherences in the $\gs_3$ representation. 
The concurrence at time $q$ of an initial pure two-qubit state under the influence of identical independent phase-damping channels
\begin{equation}C^{\footnotesize{\textrm{(PD)}}}(q;C_0,r,s)=
\max\left\{0,-\tilde{q}\,s+\sqrt{\tilde{q}^2\,r^2+(1-\tilde{q})\,C_0^2}\right\}\end{equation}
with $\tilde{q}=2q-q^2$, $s+r=2|\psi_{00}\psi_{11}|$ and $s-r= 2|\psi_{01}\psi_{10}|$, allows us to show that ESD is generic: only a zero-measure set of entangled initial states do not disentangle in finite time, those with at least one vanishing component $\psi_{ij}$. The initial states $(\ket{01}+e^{i\phi}\ket{10})/\sqrt{2}$ and $(\ket{00}+e^{i\phi}\ket{11})/\sqrt{2}$ have the maximum concurrence at time $q$, $(1-q)^2$. The probability distribution of concurrence $\tilde{p}^{\footnotesize{\textrm{(PD)}}} (C;q)
= \int d\theta\, ds\, dr\, P_J(r,s)\delta(C_q-C^{\footnotesize{\textrm{(PD)}}}(q;C_0,r,s))/2\pi$, depends on the joint probability $P_J\left(s,r\right)$, defined before. After recasting $\tilde{p}^{\footnotesize{\textrm{(PD)}}} (C;q)$ as a $2d$ integral with complicated integrands and integration regions, it is evaluated numerically. 
The probability distribution of concurrence for several values of time $q$ is shown in the inset of Fig. \ref{fig:PD}. 

\section{Mixed states.}\label{sec:mixedstates}
\begin{figure}[htpb!]
 \centering
 \includegraphics[angle=-90,width=9cm]{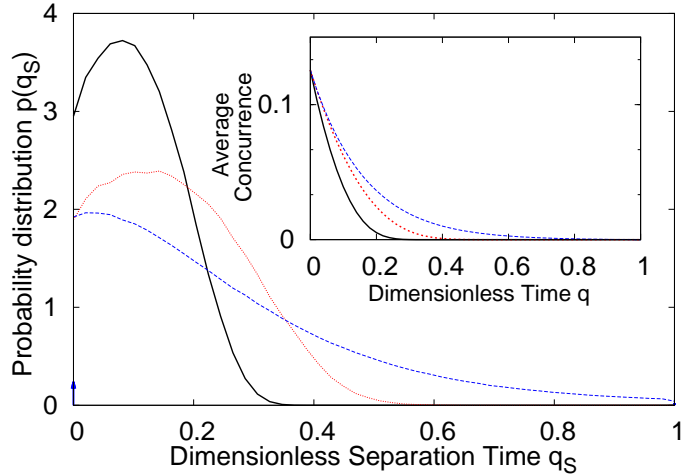}
 \caption{(Color online) Disentanglement-time probability distributions and average concurrence (Inset): solid line (black) for local depolarizing channels (D); dashed line (blue) for local amplitude damping channels (AD); and dotted line (red) for local phase damping channels (PD). The delta function at $q=0$ for the ESD time distribution is common to all cases.}
 \label{fig:MixedESD}
\end{figure}

Diferent experimental conditions call for different sets of states to be considered, in particular mixed states. Albeit no natural uniform distribution exists for them, Hilbert-Schmidt and Bures measures are commonly used. Analytical results can not be obtained along the lines presented in the main text, but it is possible to perform numerical simulations. Employing the first measure, around 24\% of the states are separable (before interaction with the channels), as signaled by a delta function contribution at time $q=0$ in Fig. \ref{fig:MixedESD}. The probability distribution of the separation times and the average concurrence show that the disentanglement process occurs first for case D (identical but independent depolarization channels), then for case PD (phase-damping channels), and finally for case AD (amplitude-damping channels), where about 2\% of the states decay asymptotically. This separation order is expected, because it depends on the behavior of the set of almost pure 
states.

\begin{figure}[htpb!]
 \centering
 \includegraphics[angle=-90,width=8.5cm]{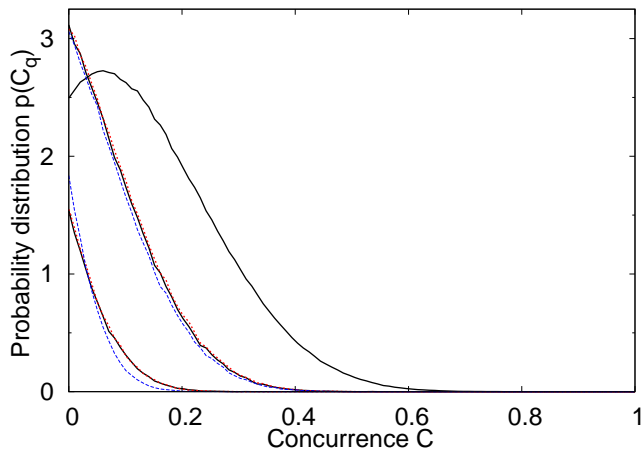}
 \caption{(Color online) Concurrence distributions: solid line (black) for local depolarizing channels (D) and times $q=0,0.15,0.31$; dashed line (blue) for local amplitude damping channels (AD) and times $q=0,0.2,0.55$; and dotted line (red) for local phase damping channels (PD) and times $q=0,0.1,0.2$. The rightmost distribution (solid) is common to all cases.}
 \label{fig:MixedProbConc}
\end{figure}

The behavior of cases D and PD, is quite similar for mixed states, up to their different time scales. In fact, not only the concurrence averages and the distributions of the disentanglement times are similar ($p^{\footnotesize{\textrm{(PD)}}}(q_S)\approx p^{\footnotesize{\textrm{(D)}}}(q_S/\alpha)/\alpha$ and $\overline{C_{q}^{\footnotesize{\textrm{(PD)}}}}\approx \overline{C_{q/\alpha}^{\footnotesize{\textrm{(D)}}}}$, where $\alpha\approx 1.56$), but also the corresponding concurrence distributions are alike. As shown in Fig. \ref{fig:MixedProbConc} the probability distributions of concurrence satisfy $\tilde{p}^{\footnotesize{\textrm{(PD)}}}(C;q)\approx \tilde{p}^{\footnotesize{\textrm{(D)}}}(C;q/\alpha)$. For early times, an appropiate time scaling shows that the concurrence probability distribution of case AD is also similar to the other two. For longer times, the relative weight of small values of concurrence is higher in this case.  

\section{A Single Noisy Channel.} \label{sec:singlechannel}
In order to have a wider spectrum of possible behaviors we also consider the  case in which only one of the qubits is affected by a dissipative channel (D, AD, PD), and the initial states are pure. The concurrence at time $q$ turns out to be $ C(q) = \max(0,x(q) C_0)$ where $x^{(D)}(q)=1-3q/2$, $x^{(AD)}(q)=\sqrt{1-q}$ and $x^{(PD)}(q)=1-q$. If we have a single depolarizing channel, the process of concurrence degradation occurs up to the reparameterized time $q=2/3$, and then all states undergo entanglement sudden death, but in the other two cases concurrence decreases steadily to zero in an asymptotic way, and at a slower pace in the case of the amplitude-damping channel. 

\section{Non-Autonomous Systems}\label{sec:nonautonomous}
We also consider non-autonomous systems \cite{Drumond2007JPA42a285308}, whose equation of motion is of the form $\dot{\rho}=\exp(-\gamma t)\mathcal{L}(\rho)$. If the Kraus operators of the usual dissipative dynamics $\dot{\rho}=\mathcal{L}(\rho)$ are $E_i(q)$, the Kraus operators of the non-autonomous version are $E_i(q/\gamma)$. That is, the dynamics at infinite time is given by $\rho(1/\gamma)$: if $\gamma>1$, all of the cases considered in this paper would suffer ESD, but if $\gamma<1$, in the cases of single or double AD and PD channels, there would be a non-zero-measure set of asymptotic entangled states. Analogously, we would have asymptotic entangled states for the single depolarizing channel if $\gamma<2/3$, and for two identical depolarizing channels if $\gamma<1-1/\sqrt{3}$. 

\section{Time-Dependent Markovian and Non-Markovian Channels}\label{sec:nonmarkovian}
\begin{figure}[htpb!]
 \centering
 \includegraphics[angle=0,width=7.8cm]{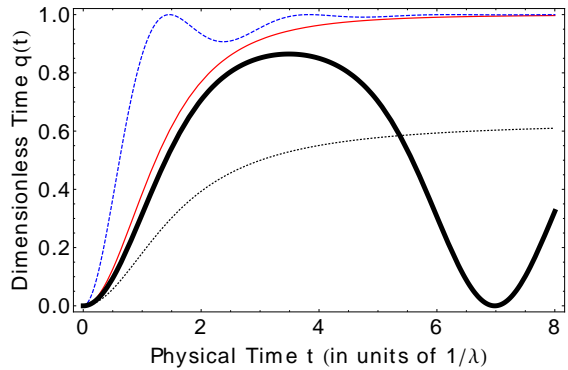}
 \caption{(Color online) Typical forms of the parameterized dimensionless time $q(t)$. From left to right: dashed line (blue) for a non-Markovian channel; solid thin line  (red) for time-dependent Markovian channel; solid thick line (black) for a non-Markovian periodic channel; and dotted line (black) for a non-Markovian time-dependent channel with incomplete decoherence for long times. Time is measured in units of $1/\lambda$, where $\lambda$ is a frequency, as described in the text.}
 \label{fig:qoft}
\end{figure}

As demonstrated in \cite{FonsecaRomero2012} an infinite number of open quantum systems can be described by the same Kraus representation used here to exemplify our results in a Markovian, constant decay-rate setting. All of our results can be mapped to any of these time-dependent Markovian or non-Markovian channels, as we show below. In order to fix ideas, we describe the interaction of a single qubit interacting with its environment by the Hamiltonian
\begin{equation}\label{eq:Hamiltonianqbreservoir}
H=\hbar\omega_0 \sigma_+\sigma_-+\sum_k \hbar \omega_k a_k^\dag a_k+(\sigma_+A
+\sigma_- A^\dag),
\end{equation}
with $A=\sum_k \hbar g_k a_k$, where $g_k$ are coupling constants and $\hbar$ is Planck's constant. We see the qubit do not need to be (strictly) a quantum memory, due to the first term of the Hamiltonian. The environment is modelled as a collection of quantum oscillators of frequency $\omega_k$, whose annihilation $a_k$ and creation $a_k^\dag$ operators satisfy the usual boson commutation relations. The transition frequency of the qubit is denoted by $\omega_0$, and
$\sigma_ \pm$ are the qubit raising and lowering operators. The initial state of the total system, $\ket{\psi}\otimes \prod_k \ket{0}_k$, is separable, with the oscillators in their ground state. For positive times, the system's state is
\begin{align*}
\label{eq:estadototal}
 \ket{\Psi(t)}=\alpha\ket{g,{0}}+\beta\left(c_0(t)\ket{e,{0}}+\sum_nc_n(t)\ket{g,1_n}\right),
\end{align*}
where $\ket{0}=\prod_k \ket{0}_k$, $\ket{1_n}=\left(\prod_{m\neq n} \ket{0}_m\right)\otimes\ket{1}_n$, and $c_0(t)$ and $c_n(t)$ are time-dependent coefficients such that $c_0(0)=1$ and $c_n(0)=0$. The Schr\"odinger equation of motion provides equations for the time derivatives of these coefficients. Formally solving the equations for the coefficients $c_n(t)$, and replacing back into the equation for $c_0(t)$, we obtain the integro-differential equation
\begin{align*}
 \dot{c}_0(t)+i\omega_0 c_0(t)+\int_0^t d\tau\, \sum_n |g_n|^2 e^{-i\omega_n(t-\tau)} c_0(\tau)=0.
\end{align*}
In the continuum limit, the environment's correlation function $\sum_n |g_n|^2 e^{-i\omega_n(t-\tau)}$ becomes the Fourier transform of the spectral density $J(\omega)$, that is, $\int_{-\infty}^\infty d\omega\,J(\omega) e^{-i\omega(t-\tau)}$. Choosing $J(\omega)=\frac{\gamma_0}{2\pi} \frac{\lambda^2}{\lambda^2+(\omega-\omega_0)^2}$, where $\lambda^{-1}$ is the environment's correlation time and $\gamma_0$ is the system's relaxation rate in the Born-Markov approximation, allows for an analytic solution for the coefficient $c_0(t)$, \cite{Breuer}
\begin{align}
c_0(t)= 
e^{-\frac{\lambda t}{2}-i \omega_0 t} \left( \cos(\Omega t)+\frac{\lambda}{2\Omega}\sin(\Omega t) \right),
\end{align} 
where $\Omega={\sqrt{\lambda  (2 \gamma_0 -\lambda )}}/{2}$. 
Tracing out the environmental degrees of freedom we obtain the qubit dynamics, which corresponds to an 
amplitude-damping (AD) channel with Kraus operators $\hat{E}_i(t) =e^{-i\omega_0 t \ket{1}\!\bra{1}} \widetilde{E}_i$, and $\widetilde{E}_0(t) = \ket{0}\!\bra{0}+\sqrt{1-q(t)}\ket{1}\!\bra{1}$ and
$\widetilde{E}_1(t) =\sqrt{q(t)}\ket{0}\!\bra{1}$. The time-dependent $q(t)=1-|c_0(t)|^2$ goes from zero at the initial time to one at large times, but it has an infinite number of intervals where it decreases from one down to a local minimum, and back to one. See the dashed line in Figure \ref{fig:qoft}, where $\gamma_0$ was chosen to be equal to $4\lambda$. Since we know the behavior of entanglement as a function of $q$, we also know it in function of $t$. In particular we see that, the intervals of decreasing $q(t)$ correspond to negative decay rates, increase of average entanglement and entanglement dispersion, and sudden birth of entanglement. If $2 \gamma_0 \leq \lambda$, contrary to what has been assumed here, the dynamics is Markovian, but the decay rate is not constant. 

It is possible to formulate time-dependent Markovian and non-Markovian models for all the noisy channels considered here (see \cite{FonsecaRomero2012}). All the details of these non-constant decay rate channels are completely captured in $q(t)$. Some typical dependencies, for a dephasing model where $q(t)=1-\exp(-\Gamma(t))$ and $\Gamma(t) = \int d\omega\,J(\omega) \frac{1-\cos\omega t}{\omega}\coth(\hbar\omega/2k_B T)$, are depicted in Fig.  \ref{fig:qoft}. Here, Bolztmann's constant is denoted by $k_B$ and the temperature of the environment by $T$. The thick solid line corresponds to a single oscillator, where $q(t)$ oscillates periodically and, hence does not have a limit for long times. The thin solid line and the dotted line correspond to an ohmic environment, for which $J(\omega)=(\omega/\lambda)\exp(-\omega/\lambda)$, for finite temperature and zero temperature, respectively. While both exemplify time-dependent Markovian processes, the latter presents incomplete decoherence, for which $q(t)$ does not 
approach one for long times.
Note that non-Markovian channels present a more varied behavior than the non-autonomous channels considered in the previous section.

\section{Conclusions and outlook.} 
This research shows that concurrence probability and separation-time distributions constitute a new, global approach to the decay of quantum entanglement, exemplified here in the case of two-qubit quantum memories under identical local decoherence mechanisms. The latter distribution follows from the former $\left(p(t_{S}) = \left. -\frac{\upd}{dt}\int_0^{C_M}\,\upd C \,\tilde{p}(C;t)\right|_{t=t_{S}}\,\right)$, because it assesses the weight of the separable states. We have observed that entanglement not only disappears suddenly but also degrades and converges to its mean in a characteristic way for different decoherence channels, when the initial states are assumed to be pure and uniformly distributed. Despite this convergence to the mean, the maximum concurrence is usually much higher than the average one. Thus, whenever the initial state can be chosen, e.g. in teleportation protocols as a quantum resource, using the most robust states offers a great advantage. 

Several variations of the main problem considered in this paper were analyzed. First, if the initial states are mixed, the global concurrence evolution is similar in all cases, except for a different time-scale, up to times where the effect of the most entangled states becomes important. The posibility of a universal behavior for initial mixed states must be further explored. Second, if the initial states are pure, but only one qubit suffers decoherence, no state separates at a finite time or all of them separate at the same time.
Finally, if the evolution of the qubits is non-Markovian or non-autonomous \cite{Drumond2007JPA42a285308}, our results must be changed. For example, it may happen that, for very long times $q(t)$ does not reach 1. Depending on the value reached, (significant) entanglement might remain for long times. In non-Markovian processes $q(t)$ might also oscillate; hence, as times goes by, entanglement not only dies suddely (degrades, concentrates) but also appears suddely (becomes enhanced, disperses). 


The derivation of entanglement statistics for interacting multiqubit systems under the action of collective or individual noisy channels, and the investigation of the consequences in quantum information processing, present an interesting challenge for future research.

\begin{acknowledgments}
This work was partially funded by Fundaci\'on para la Promoci\'on de la Investigaci\'on y la Tecnolog\'\i a (Colombia) and by by Divisi\'on de Investigaci\'on Sede Bogot\'a - Universidad Nacional de Colombia under project 10971. 
K.M.F.R. gratefully acknowledges long discussions with H.~Vinck, and helpful suggestions of B.~Garraway and R.R. Rey-Gonz\'alez.
\end{acknowledgments}

\end{document}